\documentclass[aps,prx,article,twocolumn,preprintnumbers,amsmath,amssymb,superscriptaddress]{revtex4-2}
\usepackage{graphicx}
\usepackage{dcolumn}
\usepackage{bm}
\usepackage[colorlinks=true, linkcolor=blue, urlcolor=blue, citecolor=blue]{hyperref}
\usepackage{amsmath}
\usepackage{physics}
\usepackage{ulem}

\newcommand{\T}{\mathrm{T}}

\def\ee{\mathrm{e}}

\newcommand{\ii}{\mathrm{i}}
\newcommand{\vv}[1]{\vb*{#1}}
\renewcommand{\expval}[1]{\langle{#1}\rangle}

\begin{document}

\title{Sampling Electronic Fock States using Determinant Quantum Monte Carlo}

\author{Shuhan Ding}
\affiliation{Department of Physics and Astronomy, Clemson University, Clemson, SC 29631, United States}
\affiliation{Department of Nuclear Science and Engineering, Massachusetts Institute of Technology, Cambridge, Massachusetts 02139, United States}
\author{Shaozhi Li}
\affiliation{Department of Physics and Astronomy, Clemson University, Clemson, SC 29631, United States}
\author{Yao Wang}%
\email{yao.wang@emory.edu}
\affiliation{Department of Physics and Astronomy, Clemson University, Clemson, SC 29631, United States}
\affiliation{Department of Chemistry, Emory University, Atlanta, GA 30322, United States}

\begin{abstract}
\textbf{Abstract:} Analog quantum simulation based on ultracold atoms in optical lattices has catalyzed significant breakthroughs in the study of quantum many-body systems. These simulations rely on the statistical sampling of electronic Fock states, which are not easily accessible in classical algorithms. In this work, we modify the determinant quantum Monte Carlo by integrating a Fock-state update mechanism alongside the auxiliary field. This method enables efficient sampling of Fock-state configurations. The Fock-state restrictive sampling scheme further enables the pre-selection of multiple ensembles at no additional computational cost, thereby broadening the scope of simulation to more general systems and models. Employing this method, we analyze static correlations of the Hubbard model up to the fourth order and achieve quantitative agreement with cold-atom experiments. The simulations of dynamical spectroscopies of the Hubbard and Kondo-lattice models further demonstrate the reliability and advantage of this method.
\end{abstract}

\maketitle

\section*{Introduction}
\vspace{-2mm}
A fundamental inquiry in modern condensed matter and quantum science is understanding the collective behavior of quantum many-body systems. Yet, accurately solving these complex systems with unbiased classical numerical methods continues to pose significant challenges. When multiple electrons or other degrees of freedom are entangled, the Hilbert space required to fully represent the relevant states of the system scales exponentially with the number of particles. This fast increase in the Hilbert space has significantly limited the application of wavefunction-based techniques, including exact diagonalization (ED) and density matrix renormalization group theory (DMRG)\,\cite{white1992density,schollwock2005density}, in solving many-body systems. Although quantum Monte Carlo methods do not suffer from this limitation\,\cite{sugiyama1986auxiliary, blankenbecler1981monte, sorella1989novel, white1989numerical, assaad2008world}, the fermion-sign problem and finite temperature hinder us from accessing the ground eigenstate of a many-body quantum system.

Quantum computing techniques provide a promising solution for quantum many-body systems\,\cite{feynman1982simulating}. In addition to gate-based universal quantum computers, manifest as the noisy intermediate-scale quantum (NISQ) machines in the near future\,\cite{preskill2018quantum,cao2019quantum}, have emerged as an alternative for modeling correlated electrons in quantum materials\,\cite{jaksch1998cold,somaroo1999quantum, georgescu2014quantum}. Among analog simulators, ultracold neutral atoms confined within optical lattices provide a versatile platform for simulating electronic wavefunctions within solid-state crystals\,\cite{bloch2005ultracold,gross2017quantum, arguello2019analogue, daley2022practical}. By utilizing two hyperfine states and exploiting the Feshbach resonance, precise control of the on-site Hubbard-like interaction $U$ is achievable\,\cite{jordens2008mott,schneider2008metallic,esslinger2010fermi}. Quantum gas microscopes, with their ability to sample many-body states at the single-site spatial resolution, facilitate statistical measurements for evaluating instantaneous spin and charge distributions\,\cite{bakr2009quantum, sherson2010single, greif2013short, cheuk2015quantum, haller2015single, parsons2015site, preiss2015quantum, boll2016spin, greif2016site,parsons2016site, cheuk2016observation, koepsell2020robust} as well as multi-point correlations encoding entanglement and topological orders\,\cite{schweigler2017experimental, hilker2017revealing, salomon2019direct, koepsell2019imaging, vijayan2020time, prufer2020experimental, zache2020extracting, koepsell2021microscopic}. With these progresses, quantum simulation techniques have enabled the simulation of strongly correlated electrons in system sizes inaccessible with exact numerical solutions, thus offering a preliminary insight into entanglement properties in models relevant to quantum materials.

Accessing higher-order correlations, which are crucial for wavefunctions with greater entanglement depth, necessitates increased sampling of Fock states in analog quantum simulators to reduce statistical errors. Furthermore, larger system sizes demand additional samples. Hence, the application of analog quantum simulators to highly entangled and sufficiently large problems is hindered by sampling inefficiency. To address this issue, machine learning-based methods have been proposed to expedite this process through advanced data analysis\,\cite{bohrdt2019classifying,bohrdt2021analyzing}. However, training an efficient machine learning model necessitates a substantial volume of data beforehand. Existing experimental measurements remain costly and do not yield adequate data for training an efficient machine-learning model. Taking the Hubbard model as an example, a typical analog simulation based on quantum gas microscopy collects $10^5$ snapshots, inadequate to train sophisticated deep-learning models.

The preparation of Fock-state samples has been successfully achieved using DMRG for zero-temperature systems\,\cite{ferris2012perfect} and through minimally entangled typical thermal states for finite temperatures\,\cite{White2009}. However, their applicability is primarily limited to quasi-one-dimensional systems and low-temperature regimes. In contrast, determinant quantum Monte Carlo (DQMC) is optimized for high-temperature ensembles and has been effective in simulations of $10\times10$ fermionic systems at intermediate and high temperatures. This compatibility with respect to temperature and system size makes DQMC an excellent candidate for generating Fock-state samples consistent with those obtained in analog quantum simulators. Nevertheless, conventional DQMC relies on stochastic sampling of Hubbard-Stratonovich fields rather than directly yielding Fock states of electrons, leading to inefficiencies in obtaining Fock-state samples.

To address these challenges, a configuration sampling method based on a conditional probability chain was recently proposed, iteratively constructing Fock states from specific auxiliary field configurations\,\cite{humeniuk2021numerically}. Retaining the DQMC framework, this method inherits certain limitations of DQMC, such as relatively constrained models and ensembles. 
Here, we propose an alternative approach by embedding the Fock state sampling process directly into the DQMC framework, establishing a unified Markov chain that alternates updates between the auxiliary fields and Fock-state configurations.
This algorithm, termed the Fock-State determinant quantum Monte Carlo (FDQMC), enables direct pre-selection of sampled Fock states, offering unprecedented flexibility across various ensembles and systems. With computational costs comparable to traditional DQMC, FDQMC provides statistical Fock-state samples efficiently, facilitating multi-point observables akin to those measured in quantum gas microscopy. This capability positions FDQMC as a powerful and precise emulator for cold-atom experiments. To demonstrate the capability of FDQMC, we investigate staggered magnetization, two-point, and higher-order correlations across various ensembles, successfully reproducing key features observed in quantum gas microscope experiments under comparable simulation conditions. Additionally, we extend this method to simulate the dynamical spectroscopies using examples of Hubbard and Kondo-lattice models. In the context of Kondo-lattice models, FDQMC directly simulates the spin-fermion interaction, leveraging its capacity for pre-selecting a fixed number of slave fermions, thus broadening its applicability to constrained quantum systems.

\vspace{-2mm}
\section*{Results}
\vspace{-2mm}
\subsection*{The Fock-State DQMC algorithm}
\vspace{-2mm}
For the Hubbard model, the determinant quantum Monte Carlo employs the Hubbard-Stratonovich decomposition to map the expectation of observables in an interacting system into a statistic average of measurements in an effective non-interacting system that couple to an auxiliary field\,\cite{sugiyama1986auxiliary, blankenbecler1981monte, sorella1989novel, white1989numerical, assaad2008world}. This decomposition is expressed as 
\begin{align}
    Z = \mathrm{Tr}[\rho] = \sum_x\mathrm{Tr}[ {\rho}_x],
\end{align}
where $Z$ is the partition function, $x$ is the Hubbard-Stratonovich field, and ${\rho}_x = \mathcal{T}e^{-\int_0^\beta \sum_{ij}c^\dagger_i H_{ij}[x(\tau)]c_j d\tau}$ is the density matrix for the effective non-interacting system, associated with an imaginary-time-dependent auxiliary field $x(\tau)$. Here, $H[x(\tau)]$ represents the Hamiltonian for the effective non-interacting system. 

\begin{figure}[!t]
    \centering
    \includegraphics[width=\columnwidth]{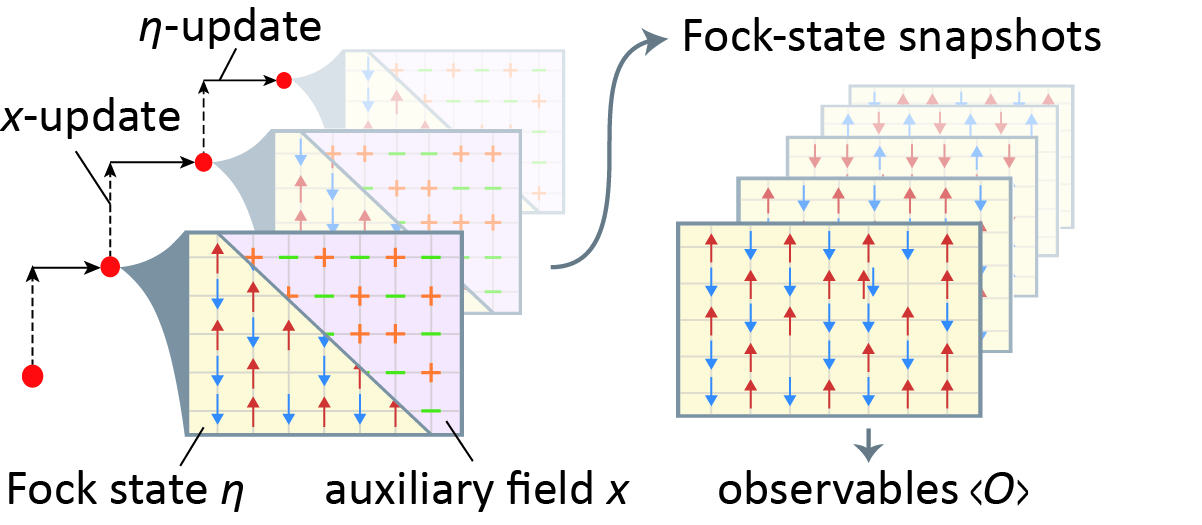}\vspace{-3mm}
    \caption{\label{fig:schematic}\textbf{Schematic illustrating the FDQMC update strategy.} The blue and red arrows represent spin-$\frac{1}{2}$ fermions, while the orange ``+" and green ``-'' symbols depict typical Ising-type auxiliary fields. Random walker of the $(x,\eta)$ pair is alternately updated in $\eta$- and $x$-directions during the Markov-chain sampling. The sampled Fock states align with the distribution of projectively measured snapshots within the system, achieved through sign reweighting. General observables are measured by statistically averaging these samples.}
\end{figure}

Unlike the traditional DQMC, we can further project ${\rho}_x$ to the Fock-state basis $\ket{\eta}$ before evaluating the expectation values of observables. That is,
\begin{align}
    Z = \sum_{x,\ket{\eta}}Z_{x,\ket{\eta}} = \sum_{x,\ket{\eta}} \mel{\eta}{\rho_x}{\eta},
\end{align}
where $\ket{\eta}$ is a binary vector that specifies a fermionic Fock state in the real-space representation. For example, $\ket{1010}$ represents the state $c_1^\dagger c_3^\dagger |0\rangle$. This projection mimics the snapshot sampling in the quantum gas microscope of quantum simulations\,\cite{gross2017quantum}. With these projections, the expectation value of an observable $O$ is calculated by
\begin{align}\label{eq: obs and Z}
    \expval{O} = Z^{-1}\sum_{x,\ket{\eta}} Z_{x,\ket{\eta}} \expval{  O}_{x,\ket{\eta}},
\end{align}
where $\expval{O}_{x,\ket{\eta}} = \mel{\eta}{O \rho_x}{\eta}/\mel{\eta}{ \rho_x}{\eta}$. Here, $Z_{x,\ket{\eta}}$ is not positive semi-definite since the projection onto a random Fock state breaks the particle-hole symmetry \cite{wu2005sign}. Therefore sign re-weighting is needed by adjusting Eq.~\eqref{eq: obs and Z} into
\begin{align}\label{eq:signAverage}
    \expval{O} = \frac{\sum_{x,\ket{\eta}} \abs{Z_{x,\ket{\eta}}}{\rm sgn}\qty(Z_{x,\ket{\eta}})\expval{  O}_{x,\ket{\eta}}}{\sum_{x, \ket{\eta}} \abs{Z_{x,\ket{\eta}}}{\rm sgn}\qty(Z_{x,\ket{\eta}})}.
\end{align}

FDQMC utilizes $\abs{Z_{x,\ket{\eta}}}$ as the joint statistical weight for $x$ and $\eta$ to generate a significant number of  $(x,\eta)$ pairs. The statistical average over these pairs allows an unbiased evaluation for the expectation value $\expval{O}$. We adopt the Metropolis-Hasting update for Markov-chain importance sampling. Specifically, for a proposed flip $(x,\eta) \rightarrow (x',\eta')$, the acceptance probability is calculated as $P_{\rm acc} = \min\{\abs{R_{\rm acc}},1\}$, where $R_{\rm acc} = Z_{x',\ket{\eta'}}/Z_{x,\ket{\eta}}$. As shown in Fig.\,\ref{fig:schematic}, FDQMC updates introduce an additional dimension compared to traditional DQMC. It achieves importance sampling of Fock states with a signed uniform weight. Ideally, for large sample size, the physical distribution of Fock-state snapshots in the original system can be reproduced by partially canceling positive-weight samples with those carrying negative weights,
\begin{align}
    \mel{\eta}{\rho}{\eta} = \sum_{x} \abs{Z_{x,\ket{\eta}}}{\rm sgn}\qty(Z_{x,\ket{\eta}}).
\end{align}
In practice, insufficient sample size may result in a non-positive distribution. However, if the sign problem is not severe, these samples can accurately reproduce all high-order correlations in an unbiased manner through Eq.~\eqref{eq:signAverage}. A systematic investigation of the sign problem is presented in Supplementary Note 2. The update and measurement strategies are detailed in \textbf{Methods}.

\begin{figure}
    \centering
    \includegraphics[width=0.8\columnwidth]{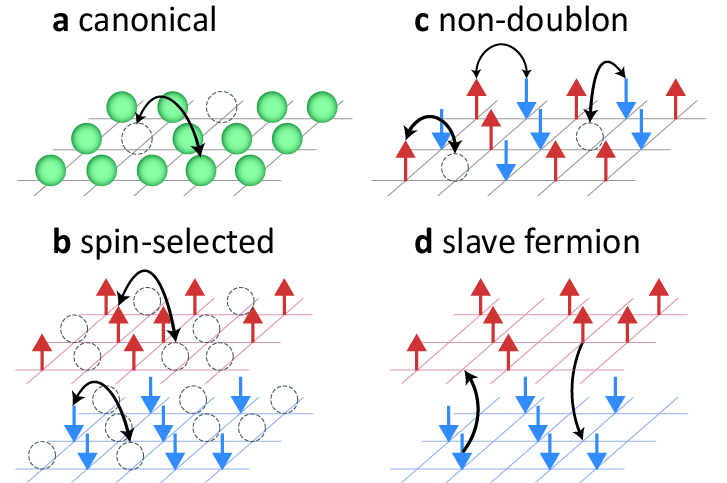}\vspace{-3mm}
    \caption{\label{fig:FRSschematic}\textbf{Ensemble selection via Fock-state restrictive updates.} \textbf{a} The canonical ensemble by swapping particles and holes. \textbf{b} The spin-selected ensemble by swapping particles and holes within each spin sector. \textbf{c} The non-doublon ensemble by swapping among singly-occupied electrons and holes. \textbf{d} Slave fermions by flipping the on-site spins.\vspace{-1mm}
    }
\end{figure}

Unlike traditional DQMC, direct pre-selection for any ensembles and physical constraints is easily achieved in FDQMC, through the Fock-state restrictive sampling (FRS). This capability is essential for accurately evaluating observables, particularly doping-sensitive observables, in finite systems where the grand-canonical and canonical ensembles differ significantly, and particle number fluctuations can introduce substantial noise. FRS scheme is analogous to the post-selection method used in analyzing quantum gas microscope experiments\,\cite{schweigler2017experimental, hilker2017revealing, salomon2019direct, koepsell2019imaging, vijayan2020time, prufer2020experimental, zache2020extracting, koepsell2021microscopic}, but is performed before the Monte Carlo update without generating unused samples. The grand-canonical ensemble is simulated by default if no restrictions are applied. In this work, we examine three different ensembles. The canonical ensemble with a fixed particle number can be simulated by randomly swapping a site-$i$ particle and a site-$j$ hole for each $\eta$-update (see Fig.\,\ref{fig:FRSschematic}\textbf{a}). Building upon this, the spin-selected ensemble further mandates a consistent total spin (in the z-direction) and is realized by restricting particle-hole swaps within each spin sector (see Fig.\,\ref{fig:FRSschematic}\textbf{b}). Finally, the non-doublon ensemble excludes double occupation on the top of the spin-selected ensemble, often used in quantum simulations. In the context of the Hubbard model, this ensemble serves as an extension of the $t$-$J$ model, including all higher-order spin-exchange processes. The non-doublon ensemble is achieved by swapping up-spins,  down-spins, and holes individually (see Fig.\,\ref{fig:FRSschematic}\textbf{c}). As will be elaborated later, the non-doublon ensemble is more effective in signaling high-order correlations. Additionally, constraints on fixing on-site fermion number enable FDQMC to simulate spin-fermion interactions, such as Kondo coupling, through slave fermions (see Fig.\,\ref{fig:FRSschematic}\textbf{d}).

\vspace{-2mm}
\subsection*{Magnetization in a Hubbard model}
\vspace{-2mm}
We apply FDQMC to the single-band Fermi-Hubbard model in a 2D square lattice, whose Hamiltonian is
\begin{align}
    \mathcal{H}_{\rm H} = -t\sum_{\expval{ij},s} (c^\dagger_{is}c_{js} + {\rm h.c.}) + U\sum_{i}n_{i\uparrow}n_{i\downarrow}\,.
\end{align}
Here, $t$ denotes the hopping between nearest neighbors, and $U > 0$ represents the on-site repulsive interaction. In this section, we examine static two-point and higher-order correlations across various ensembles.

\begin{figure}[!b]
    \centering
    \includegraphics[width=\linewidth]{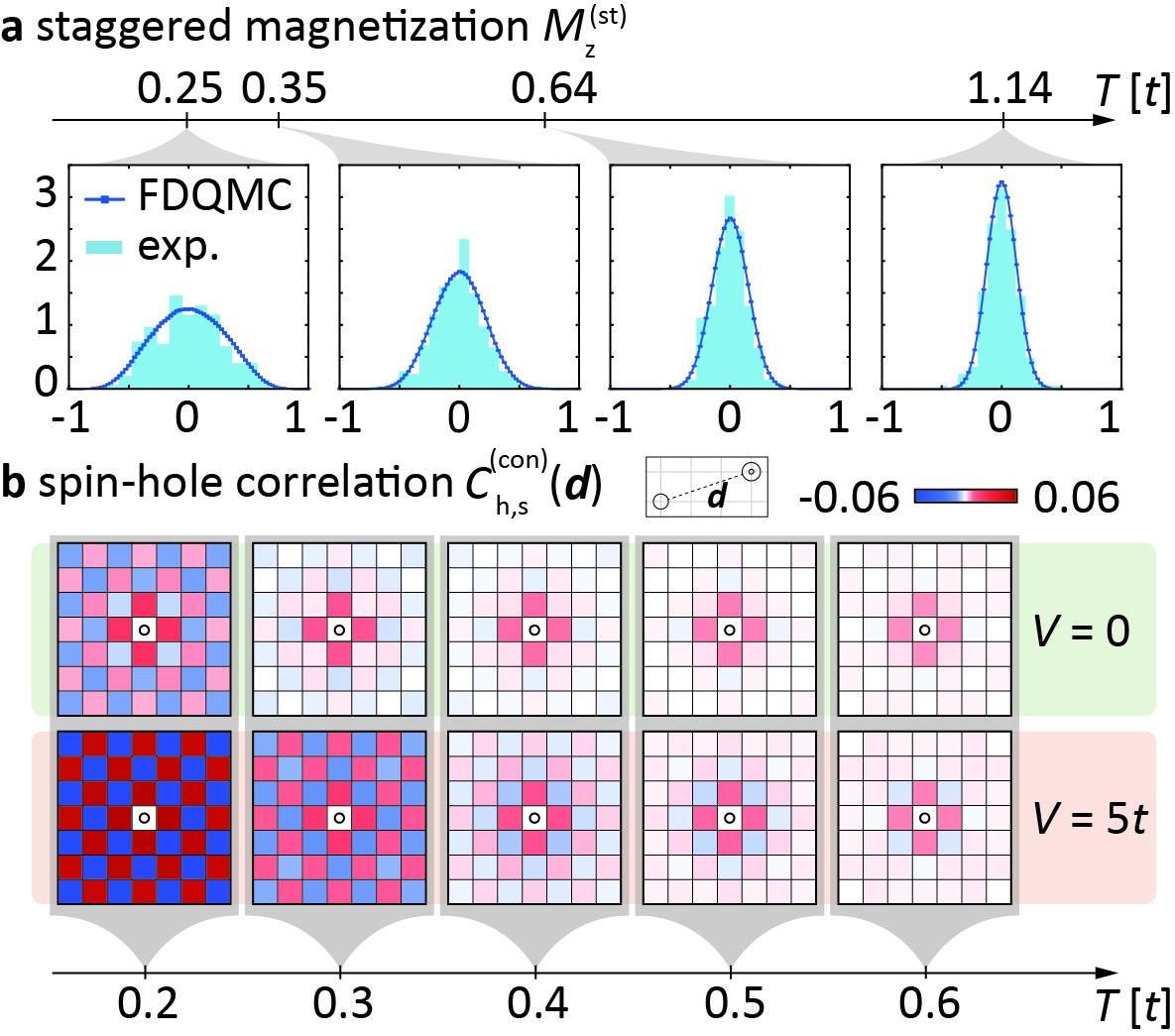}\vspace{-3mm}
    \caption{\label{fig:twoPointCorr}\textbf{Magnetization in undoped and single-hole-doped Hubbard models.} \textbf{a} Distributions of the staggered magnetization $M^{\rm (st)}_z$ in the undoped Hubbard model, calculated using Fock-state samples generated by FDQMC at different temperatures $T$. The standard error is smaller than the data points. The bars represent results obtained from cold-atom experiments in Ref.~\onlinecite{mazurenko2017cold}. \textbf{b} The spatial distribution of $C_{\rm h,s}^{\rm (con)}(\vv d)$ calculated within the non-doublon ensemble at different temperatures $T$, where the upper (lower) panel corresponds to pinning potential $V=0$ ($V=5t$).}
\end{figure}

Leveraging the capability to select specific ensembles, the simulated results of FDQMC can be benchmarked against the quantum gas microscope experiments in Ref.~\onlinecite{mazurenko2017cold}. Figure\,\ref{fig:twoPointCorr}\textbf{a} shows the distributions of staggered magnetization at half-filling, defined as 
\begin{align}
    M^{\rm (st)}_z = \frac{2}{N}\sum_{\vv i}(-1)^{i_x+i_y}S^{\rm z}_{\vv i}.
\end{align}
The simulated distributions are statistically derived from $\sim 4\times 10^6$ Fock-state samples by FDQMC, while the experimental results are obtained using $\sim$ 250 snapshots via quantum gas microscope\,\cite{mazurenko2017cold}. Here, $S_{\vv i}^{\rm z} = ({n}_{\vv i\uparrow} - {n}_{\vv i\downarrow})/2$ represents the $z$-component of the spin at site $\vv i$, $N$ denotes the total number of sites, and the Hubbard interaction is set to $U = 7.2\,t$ in align with experiments. We select an $8\times10$ periodic square lattice and employ the canonical ensemble to compare with the experimental results measured within a circular central region containing approximately 80 sites. At high temperatures, the simulated $M^{\rm (st)}_z$ closely aligns with the distribution obtained from experiments, appearing as a Gaussian envelope center at zero with spin symmetry. As the temperature falls below the spin-exchange energy $J\sim 0.55\,t$, the distribution noticeably broadens due to quantum fluctuations driven by antiferromagnetic (AFM) correlations \cite{humeniuk2017full}. These fluctuations are reflected numerically by more distinct Fock-state configurations within the same sample volume. Therefore, the experimental distribution starts to deviate from a smooth and symmetric distribution, constrained by its 250 snapshots. In contrast, the distribution obtained by FDQMC can reach exact solutions with minimal statistical error, benefiting from the extensive sample volume ($\sim 4\times 10^6$). For a detailed comparisons, please refer to Supplementary Note 1.

When a single hole is introduced into the AFM background, it can influence magnetization\,\cite{sachdev1989hole,martinez1991spin, dagotto1992static, bala1995spin}. Here, we examine a single-hole-doped Hubbard model with $U = 8t$ on a $6\times6$ periodic square lattice. Given the odd number of electrons with the presence of a single hole, we opt for the spin-selected and non-doublon ensembles and set the total spin to be $1/2$. With this selected orientation, the magnon dressing of the spin polaron can be visualized through the (connected) spin-hole correlation\,\cite{blomquist2020unbiased,koepsell2019imaging}
\begin{align}
    C^{\rm (con)}_{\rm h,s}(\vv d) = 2\expval{{n}_{\vv 0}^{\rm h} {S}_{\vv d}^{\rm z}}/\expval{{n}_{\vv 0}^{\rm h}} - 2\sum_{\vv r}\expval{S_{\vv r}^{\rm z}}/N,
\end{align}
where ${n}_{\vv 0}^h = (1- {n}_{\vv 0\uparrow})(1- {n}_{\vv 0\downarrow})$ is the hole operator at the origin. As shown in Fig.\,\ref{fig:twoPointCorr}\textbf{b}, the high-temperature system is disordered, with no magnetization except that the additional spin moment accumulates near the nearest neighbor of the hole. With the decrease of temperature, the AFM order starts to develop, and the motion of the hole is dressed by the disturbance of the AFM correlations, resulting in the checkerboard distribution of $C^{\rm (con)}_{\rm h,s}(\vv d)$ throughout the system.

The magnetization is further influenced by the mobility of the hole. We delve into this effect by adding a pinning potential $V$ at the origin site to tune the mobility of the dopant. This leads to the modified Hamiltonian
\begin{align}
    H_{V} = H_{\rm H} +  V(n_{\vv 0\uparrow} + n_{\vv 0\downarrow})\,.
\end{align}
The pining potential can be realized experimentally using an optical tweezer in an optical lattice\,\cite{zhang2006manipulation,beugnon2007two, koepsell2019imaging}. While similar to the $V = 0$ results at high temperatures, the staggered magnetization starts to develop at a higher temperature with a strong pinning potential. In the lowest temperature ($T=0.2t$), the magnetization resembles a N\'{e}el state with a missing down-spin at the origin. Given the reduced hole's mobility with a pronounced $V$, it serves as a geometric defect at low temperatures. 

\begin{figure}[!t]
    \centering
    \includegraphics[width=\linewidth]{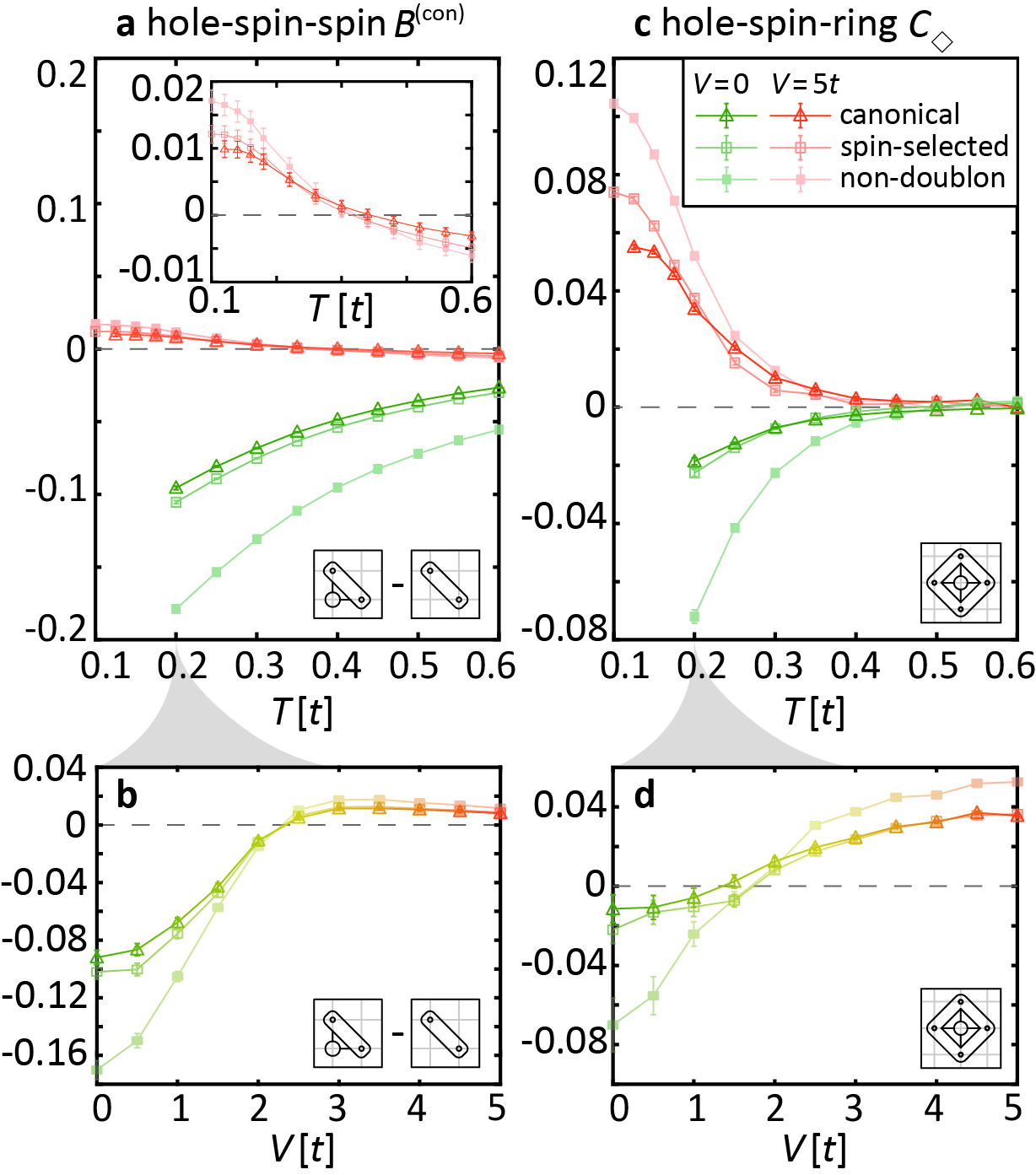}\vspace{-3mm}
    \caption{\label{fig:higherOrderCorr} \textbf{High-order correlations of the Hubbard model with a single hole.} \textbf{a} Temperature dependence of third-order correlations $B^{\rm(con)}(\vv 0, \hat{\vv x}, \hat{\vv y})$ for $V=0$ (green) and $V=5t$ (orange), simulated in the canonical (triangle), spin-selected (open square), and non-doublon (solid square) ensembles. The inset shows the same data with the scaled axis. \textbf{b} The dependence on the pinning potential $V$ at $T=0.2\,t$. \textbf{c},\textbf{d} Same as \textbf{a},\textbf{b} but for the fifth-order $C_\diamondsuit$. The cartoon in each panel illustrates the geometry of each correlation. Error bars represent the standard error of the Monte Carlo data.
    \vspace{-2mm}
    }
\end{figure}

\vspace{-2mm}
\subsection*{High-order correlations}
\vspace{-2mm}
A significant advance of quantum simulations is the analysis of multi-point high-order correlations, which extends beyond the capabilities of conventional spectroscopic measurements in solid-state materials. In the context of the Hubbard model, connected high-order correlations have been utilized to uncover the hidden orders, polaronic wavefunctions, and entanglement\,\cite{schweigler2017experimental, hilker2017revealing, salomon2019direct, koepsell2019imaging, vijayan2020time, prufer2020experimental, zache2020extracting, koepsell2021microscopic, humeniuk2021numerically}. 

Following the formalism in Ref.~\cite{wang2021higher}, we examine the property of a single-hole-doped Hubbard model using FDQMC on an $6\times6$ square lattice through the analysis of the third- and fifth-order correlations. The connected part of the third-order correlation is defined as\,\cite{koepsell2019imaging, wang2021higher}
\begin{align}
    B^{\rm (con)}(\vv r, \vv r', \vv r'') = 4\expval{n_{\vv r}^h S_{\vv r'}^{\rm z} S_{\vv r''}^{\rm z}}/\expval{n_{\vv r}^h} - 4\expval{S_{\vv r'}^{\rm z} S_{\vv r''}^{\rm z}},
\end{align}
highlighting the impact of a hole (at site $\vv r$) on the spin-spin correlation $\expval{S_{\vv r'}^{\rm z} S_{\vv r''}^{\rm z}}$ (see the inset of Fig.~\ref{fig:higherOrderCorr}\textbf{a}). The temperature dependence of $B^{\rm(con)}(\vv r, \vv r',\vv r'')$, for the diagonal spin correlations $(\vv r',\vv r'') = (\hat{\vv x}, \hat{\vv y})$ with respect to the hole at $\vv r = 0$, is shown in Fig.\,\ref{fig:higherOrderCorr}\textbf{a}. Its disconnected part, i.e.~$4\expval{S_{\vv r'}^{\rm z} S_{\vv r''}^{\rm z}}$, is positive in the AFM background, and $B^{\rm(con)}(\vv r, \vv r',\vv r'')$ indicates the enhancement or diminishment of this correlation near a hole. Without the pinning potential, $B^{\rm (con)}$ consistently exhibits a negative value, reflecting the disturbance on the AFM spin correlations by the hole's motion. This connected correlation serves as a fingerprint for a spin polaron. Nonetheless, with the introduction of a strong pinning potential $V = 5t$, the sign of $B^{\rm (con)}$ transitions to positive at sufficiently low temperature ($T<0.25\,t$). Such a flip signals an ``anti-screening'' effect of the hole at low $T$, strengthening the spin correlations near the hole. This effect is attributed to the reduction of spin fluctuations with fewer neighboring sites, when the hole is immobile and becomes effectively a geometric defect\,\cite{wang2021higher}. Such a transition from polaronic screening to anti-screening only occurs when $V$ is adequately large to surpass the kinetic energy of the hole (see Fig.~\ref{fig:higherOrderCorr}\textbf{b}). It is important to note that, although the results from different ensembles vary quantitatively, the critical temperature and pinning potential for the transition remain unaffected by the choice of ensemble. 

Another high-order correlation depicting the single-hole dynamics is the fifth-order hole-spin-ring correlation, defined as\,\cite{bohrdt2021dominant, wang2021higher}
\begin{align}
    C_{\rm \diamondsuit} = 2^4\expval{n_{\vv 0}^{\rm h} {S}_{\vv r+ \hat{\vv x}}^{\rm z} {S}_{\vv r+ \hat{\vv y}}^{\rm z} {S}_{\vv r - \hat{\vv x}}^{\rm z} {S}_{\vv r - \hat{\vv y}}^{\rm z}}/\expval{n^{\rm h}_{\vv 0}}\,,
\end{align}
where the hole at the origin is encircled by a spin ring. Similar to the third-order correlation, $C_{\rm \diamondsuit}$ remains negative for a mobile hole across all temperatures (see Fig.~\ref{fig:higherOrderCorr}\textbf{c}), reflecting the string excitation caused by the formation of spin polaron\,\cite{bohrdt2021dominant}. This negativity stems from spin correlations of the AFM background, hence intensifying at lower temperatures. With a pinning potential in place, this fifth-order correlation aligns with the $V=0$ scenario at high temperatures due to the lack of AFM order. Yet, as the temperature drops significantly below $J$, the emergence of AFM correlation and the suppression of quantum fluctuations by the geometric defect help the development of a pronounced spin-ring correlation surrounding the immobile hole. This effect is evidenced by the substantial positive values of $C_{\rm \diamondsuit}$ at low temperatures ($T<0.3\,t$). The transition from negative to positive also occurs around $V_c\sim 2t$ (see Fig.~\ref{fig:higherOrderCorr}\textbf{d}), indicating that the coincident underlying anti-screening physics as observed in the third-order correlation $B^{\rm (con)}$.

\begin{figure}[t]
    \centering
    \includegraphics[width=\columnwidth]{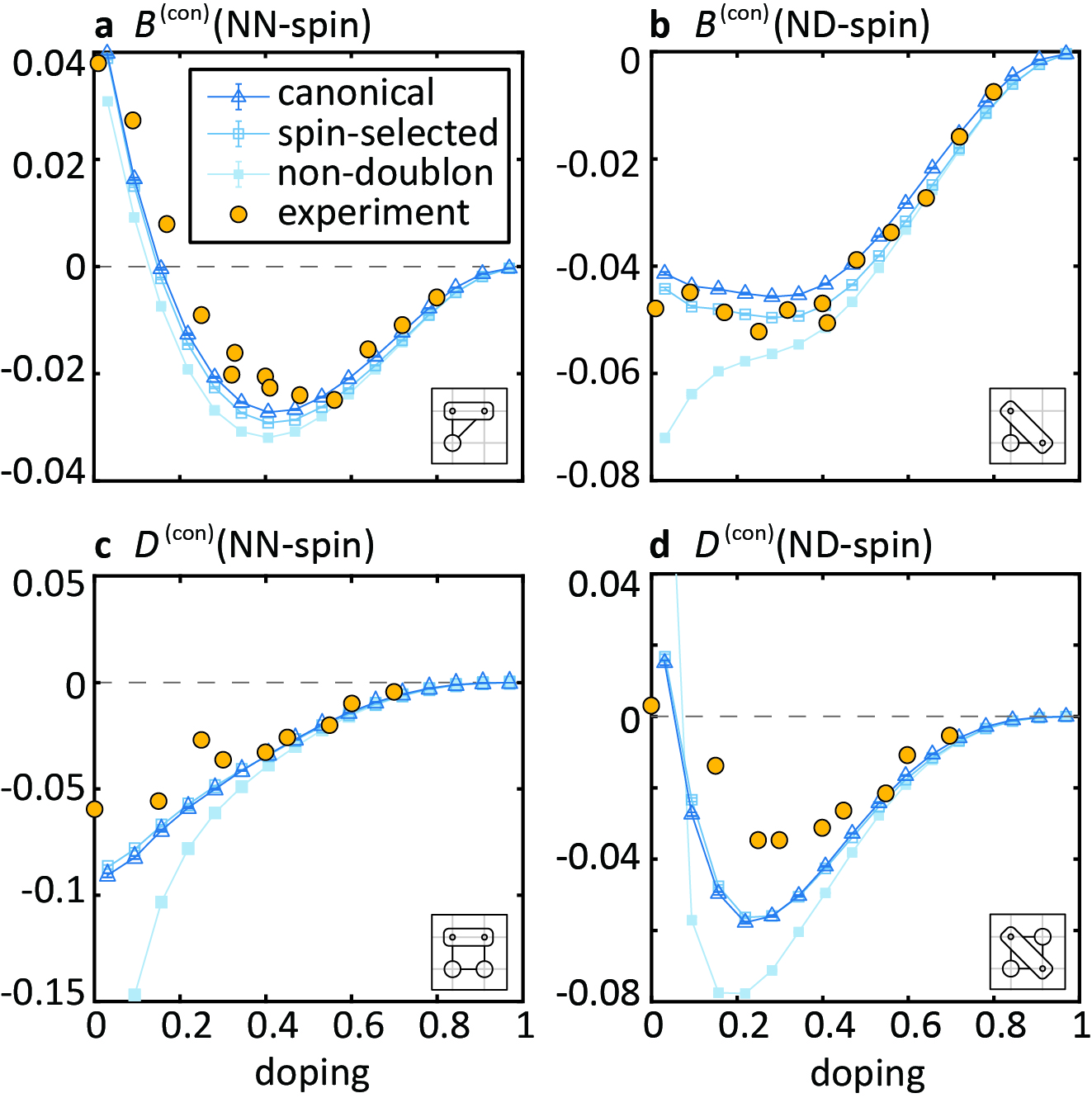}\vspace{-3mm}
    \caption{\label{fig:doping}\textbf{High-order correlations in doped Hubbard models.} \textbf{a},\textbf{b} Doping dependence of the connected third-order correlation $B^{\rm(con)}$ for \textbf{a} nearest-neighbor (NN) and \textbf{b} nearest-diagonal (ND) spins, evaluated within the canonical (triangle), spin-selected (open square), and non-doublon ensembles (solid square). The simulations are conducted on an $8\times8$ lattice with temperature $T=0.5\,t$ and Hubbard interaction $U=8\,t$. Yellow dots represent results from cold-atom experiments in Ref.~\onlinecite{koepsell2021microscopic} with similar parameters. \textbf{c},\textbf{d} Doping dependence of the connected fourth-order correlation $D^{\rm(con)}$ for \textbf{c} NN and \textbf{d} ND configurations, simulated under identical conditions to \textbf{a} and \textbf{b}. The cartoon in each panel illustrates the geometry of the correlation. All the observables are normalized by a factor of $[1/\sigma(2S^{\rm z}_{\vv r})]^2$. Error bars represent the standard error of the Monte Carlo data.
    }
\end{figure}

Expanding our analysis to higher dopings, the fermion-sign problem becomes more pronounced at low temperatures, restricting our simulations to relatively high temperatures. Here, we choose $U = 8t$ and $T = 0.5t$ on the periodic $8\times8$ lattice, matching the conditions of cold-atom experiments in Ref.~\onlinecite{koepsell2021microscopic}, where $U = 7.4(8)t$ and $T = 0.52(5)t$. An even number of doped holes are used for the FDQMC simulation to ensure that the total spin of the spin-selected and non-doublon ensembles is zero. This setting reflects the experimental reality and simplifies the simulation, as all low-order correlations involving an odd number of spin operators are nullified.

Extending the simulation of the connected third-order correlation $B^{\rm (con)}(\vv r, \vv r+\vv r', \vv r+\vv r'')$ into finite doping, we focus on two key distances indicative of spin polaron wavefunction: the nearest-neighbor-spin $B^{\rm (con)}(\mathrm{NN\!-\!spin})=B^{\rm (con)}(\vv r, \vv r+\hat{\vv y}, \vv r+\hat{\vv x}+\hat{\vv y})$ and the nearest-diagonal-spin $B^{\rm (con)}(\mathrm{ND\!-\!spin})=B^{\rm (con)}(\vv r, \vv r+\hat{\vv x}, \vv r+\hat{\vv x}+ \hat{\vv y})$ (illustrated in the insets of Figs.~\,\ref{fig:doping}\textbf{a} and \textbf{b}). Without the pinning potential, the system is translational symmetric, and the choice of $\vv r$ is irrelevant. As shown in Fig.\,\ref{fig:doping}\textbf{a}, $B^{\rm (con)}(\mathrm{NN\!-\!spin})$ obtained from all ensembles exhibits a rapidly decrease from a significantly positive value, transitioning to negative at $\sim18\%$ doping --- a change potentially linked to the temperature-independent quasi-particle interruption observed in ARPES studies of cuprates\,\cite{chen2019incoherent}. This sign flip reflects the breakdown of spin polaron with increasing doping. When comparing results from the canonical ensemble with cold atom experiments in Ref.\cite{koepsell2021microscopic}, a consistent agreement is observed throughout all dopings. This consistency further validates the efficacy of FDQMC samples in mirroring quantum simulation snapshots. A similar agreement is observed for $B^{\rm (con)}(\mathrm{ND\!-\!spin})$, as presented in Fig.\,\ref{fig:doping}\textbf{b}. Results from the non-doublon ensemble, however, deviate from the canonical ensemble and experimental results in the low doping regime, attributed to the exclusion of doublon-hole fluctuations.

The non-monotonic doping dependence of $B^{\rm (con)}(\mathrm{NN\!-\!spin})$ indicates that the doped Hubbard model maintains a strongly correlated state beyond the breakdown of the spin polaron, which has also been suggested by the persistent spin fluctuations observed in cuprates\,\cite{dean2013persistence, le2011intense,ishii2014high}. Particularly, potential interactions between holes may be mediated by the overlap of two spin-polarons\,\cite{schrieffer1989dynamic}. Therefore, we examine the fourth-order correlations involving two holes\,\cite{koepsell2021microscopic}
\begin{equation}
\begin{aligned}
    &\quad D^{\rm (con)}(\vv r_1, \vv r_2, \vv r_3, \vv r_4)\\
    &= \frac{1}{\expval{n^{\rm h}_{\vv r_1}\mkern-2mun^{\rm h}_{\vv r_2}}} \mkern-2mu\Big[\expval{n^{\rm h}_{\vv r_1}\mkern-2mun^{\rm h}_{\vv r_2}\mkern-2muS^{\rm z}_{\vv r_3}\mkern-2muS^{\rm z}_{\vv r_4}}-\expval{n^{\rm h}_{\vv r_1}} \expval{n^{\rm h}_{\vv r_2}S^{\rm z}_{\vv r_3}S^{\rm z}_{\vv r_4}}-\expval{n^{\rm h}_{\vv r_2}} \\
    &  \cdot\mkern-2mu  \expval{n^{\rm h}_{\vv r_1}\mkern-2muS^{\rm z}_{\vv r_3}\mkern-2muS^{\rm z}_{\vv r_4}}\mkern-2mu-\mkern-2mu\expval{\mkern-1mun^{\rm h}_{\vv r_1}\mkern-2mun^{\rm h}_{\vv r_2}\mkern-1mu}\expval{S^{\rm z}_{\vv r_3}\mkern-2muS^{\rm z}_{\vv r_4}}
    \mkern-2mu+\mkern-2mu 2\expval{\mkern-1mun^{\rm h}_{\vv r_1}\mkern-1mu}\expval{\mkern-1mun^{\rm h}_{\vv r_2}\mkern-1mu}\mkern-2mu\expval{S^{\rm z}_{\vv r_3}\mkern-2muS^{\rm z}_{\vv r_4}}\mkern-2mu\Big]\,,
\end{aligned}
\end{equation}
This connected correlation quantifies the net effect of a pair of holes on adjacent spin correlations, compared to separated ones. When the examined two spins and two holes form a plaquette, i.e.~NN-spin and ND-spin as shown in Figs.~\ref{fig:doping}\textbf{c} and \ref{fig:doping}\textbf{d}, $D^{\rm (con)}$ manifests significant values across a wide range of doping (up to $\sim60$\%). In particular, $D^{\rm (con)}(\mathrm{NN\!-\!spin})$ is consistently negative, reflecting a tendency for spin polarons to share spin defects. This correlation monotonically decreases as the system is doped away from the AFM phase. At the same time, $D^{\rm (con)}(\mathrm{ND\!-\!spin})$ becomes significantly negative only near 20\% doping, where the spin polarons break down. Both the NN- and the ND- $D^{\rm (con)}$ suggest the preference of spin-singlet around the closest proximity of the hole pair, consistent with experimental findings\,\cite{koepsell2021microscopic}.

When compared against various theories, it has been found that analytical wavefunctions ansatzes fail to capture the doping evolution observed in experiments\,\cite{koepsell2021microscopic}. Numerical simulations using finite-temperature ED have successfully reproduced third-order correlations $B^{\rm (con)}$ and qualitatively traced the trend of the fourth-order correlations $D^{\rm (con)}$. However, discrepancies of a factor of 2 to 2.5 are present in the ED simulations, largely due to the finite-size effects. Using the FDQMC method at the same size and temperature as the experiments, we manage to closely match experimental results for $D^{\rm (con)}(\mathrm{NN\!-\!spin})$ and significantly reduce the discrepancy of $D^{\rm (con)}(\mathrm{ND\!-\!spin})$ to around $50$\%. Since the DQMC simulations are unbiased at this temperature and system size, the remaining mismatch likely stems from the inhomogeneity and the uncertainty of model parameters in experiments or the distinction in boundary conditions. Upon comparing across different ensembles, we find that high-order correlations evaluated in the canonical and the spin-selected ensemble are similar, whereas the non-doublon ensemble leads to more pronounced correlations below 40\% doping. This indicates that the non-doublon ensemble is more suitable for elucidating genuine hole-spin correlations and uncovering their entanglement, due to the exclusion of irrelevant Fock states that involve double occupation. 

\subsection*{Dynamical correlations and spectroscopies}
\vspace{-2mm}
While simulating dynamical correlations presents challenges with analog quantum simulators\,\cite{knap2013probing}, the Fock state sampling can be extended to the analysis of unequal-time correlations and allows for the emulation of spectroscopies similar to traditional DQMC. For a specific configuration of $(x,\eta)$, the unequal-time Green's function is calculated as (assuming $\tau_1 \ge \tau_2$)
\begin{eqnarray}
    \expval{c_i(\tau_1)c_j^\dagger(\tau_2)}_{x,\ket{\eta}} &=& \qty[B_x(\tau_1,\tau_2)G_{x,\ket{\eta}}(\tau_2,\tau_2)]_{ij},\\
    \expval{c_j^\dagger(\tau_1)c_i(\tau_2)}_{x,\ket{\eta}} &=& \qty[\big(I - G_{x,\ket{\eta}}(\tau_2,\tau_2)\big)B^{-1}_x(\tau_1,\tau_2)]_{ij}\nonumber
\end{eqnarray}
using the numerically stable method introduced in Ref.~\onlinecite{feldbacher2001efficient}. Other dynamical observables are derived from Green's functions using Wick's theorem. Here, we discuss two representative examples: The single-particle spectrum $A(\vv k,\omega)$ measures the evolution of an individual electron, and the dynamical spin structure factor $S(\vv q, \omega)$ measures the propagation of a spin excitation (see definitions in \textbf{Methods}). Both $A(\vv k,\omega)$ and $S(\vv q, \omega)$ are analytically continued using the maximum entropy method\,\cite{jarrell1996bayesian}. For the sake of visualization, we normalize $S(\vv q, \omega)$ for each momentum, denoted as $\tilde{S}(\vv q, \omega)$, while the original data are shown in Supplementary Note 3.

\begin{figure}[t]
    \centering
    \includegraphics[width=\linewidth]{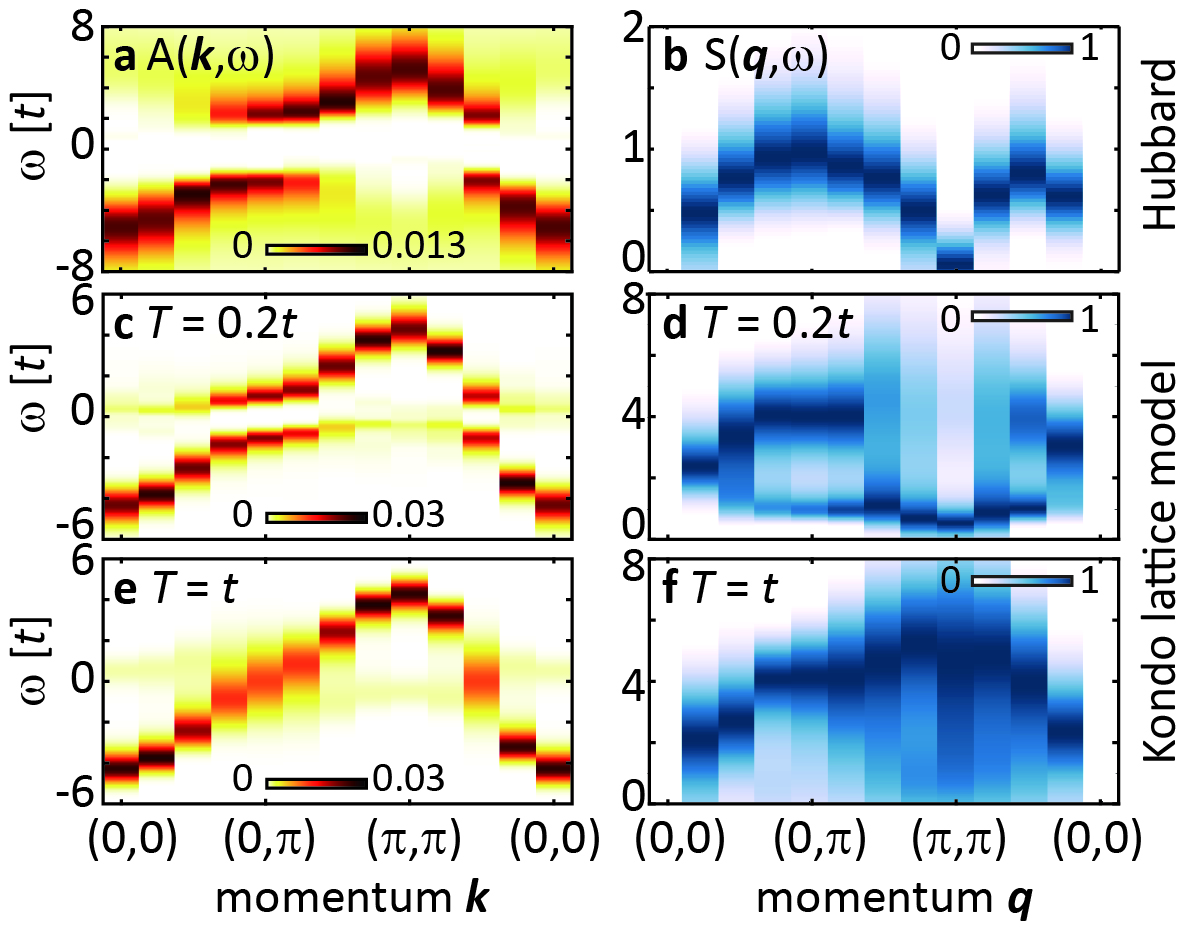}\vspace{-2mm}
    \caption{\label{fig:spectrum} \textbf{Dynamical spectroscopies simulated using FDQMC.} \textbf{a} The single-particle spectrum $A(\vv k, \omega)$ and \textbf{b} normalized spin-structure factor $\tilde{S}(\vv q,\omega)$ for a half-filled Hubbard model on an $8\times8$ cluster with Hubbard interaction $U=8\,t$ and temperature $T=0.2\,t$. \textbf{c}-\textbf{f} Same as \textbf{a} and \textbf{b} but for the half-filled Kondo lattice model on an $8\times8$ cluster with Kondo interaction $J_{\rm K} = 2\,t$. The spectra in \textbf{c} and \textbf{d} are obtained at a low temperature ($T=0.2\,t$), while those in \textbf{e} and \textbf{f} correspond to higher temperature ($T=t$) conditions. }
\end{figure}

Building on above analyses, we first study the half-filled Hubbard, utilizing an $8\times8$ cluster within the canonical ensemble. As shown in Fig.~\ref{fig:spectrum}\textbf{a}, a Mott gap of $\sim4t$ is evident in $A(\vv k,\omega)$. The Hubbard interaction also leads to the formation of 2D AFM order, as manifested by the magnon dispersion in $\tilde{S}(\vv q, \omega)$ (see Fig.~\ref{fig:spectrum}\textbf{b}). Different from small-cluster ED simulations, the $\tilde{S}(\vv q, \omega)$ evaluated from FDQMC correctly portrays the splitting between the nodal $(\pi/2,\pi/2)$ and the anti-nodal $(0,\pi)$ magnons, a discrepancy stemming from higher-order spin-exchange processes at large systems.

To demonstrate the advantage of the FRS scheme in FDQMC, we turn our attention to the Kondo lattice model, described by the Hamiltonian
\begin{equation}\label{eq:HamKondo}
    \mathcal{H}_{\rm K} \mkern-2mu=\mkern-2mu -t\mkern-4mu\sum_{\expval{ij},s} \mkern-2mu(c^\dagger_{is}c_{js} + {\rm h.c.}) + \frac{J_{\rm K}}{2}\mkern-4mu\sum_{i,s,s'}\mkern-2muc^\dagger_{is}\vv \sigma_{ss'} c_{is'}\mkern-2mu\vdot\mkern-2mu\vv S^f_{i}\,.
\end{equation}
Here, the spin of an itinerant electron couples to a localized spin-1/2 moment $\vv S^f_{i}$ with strength $J_{\rm K}$. By decomposing $\vv S^f_{i}$ using the slave-fermion representation  $(1/2)\sum_{s,s'}f^\dagger_{is}\vv \sigma_{ss'} f_{is'}$, the Kondo coupling in Eq.~\eqref{eq:HamKondo} is equivalent to an effective local, fermionic interaction
\begin{align}\label{eq:effKondoInt}
	- \frac{J_{\rm K}}{4}\qty(\sum_{s}c^\dagger_{is}f_{is} + {\rm h.c.})^2 + {\rm const.}\,,
\end{align}
which is feasible for simulation using DQMC algorithms\,\cite{capponi2001spin}. However, the slave-fermion decomposition requires the constraint $n^f_{i\uparrow} + n^f_{i\downarrow} = 1$, a condition not inherently met by traditional DQMC at finite temperatures. Common solutions to this dilemma include switching to the Anderson model with $f$-site repulsion $U_f\propto 1/{J_{\rm K}}$, or introducing a strong $U_f$ to suppress both double and zero occupancies\,\cite{capponi2001spin}. These approaches, however, induce approximations and potential bias. The strong artificial repulsion also affects the numerical stability, resulting in severe sign problems when doping. In contrast, the FRS scheme of  FDQMC strictly adheres to the slave-fermion constraint, thereby serving as an exact finite-temperature solver for Kondo-type spin-fermion interactions.

Figures \ref{fig:spectrum}\textbf{c}-\textbf{f} present FDQMC-simulated spectral results of a half-filled Kondo-lattice model in a $8\times8$ cluster. We choose the Kondo coupling ${J_{\rm K}} = 2t$ and consider itinerant electrons in the canonical ensemble. The Kondo resonance occurs at a low temperature $T = 0.2t$, resulting in the hybridization gap and the heavy-fermion dispersion in $A(\vv k,\omega)$ (see Fig.~\ref{fig:spectrum}\textbf{c}). At the same time, $\tilde{S}(\vv q, \omega)$ exhibits pronounced magnon dispersion at $T = 0.2\,t$ (see Fig.~\ref{fig:spectrum}\textbf{d}). The excitation energy at $\vv q = (\pi,\pi)$ remains finite, revealing the spin-gapped phase for $J > J_c \approx 1.45 t$\,\cite{capponi2001spin}. At high temperatures, e.g.~$T = t$ presented in Figs.~\ref{fig:spectrum}\textbf{e} and \textbf{f}, the hybridization gap closes, resulting in diminished magnon excitations. The decoupling between itinerant electrons and local spins reflects the asymptotic freedom of the Kondo lattice ($J_{\rm K}>0$) in the high-temperature limit.

\section*{Discussion}
\vspace{-2mm}
The FDQMC algorithm enables importance sampling of the joint distribution of Fock-state samples and auxiliary fields. Since the $\eta$-update is independent of the specific type of the auxiliary field, this approach is adaptable across various DQMC and auxiliary field QMC algorithms. As shown in Supplementary Note 4, incorporating Fock-state sampling does not significantly influence the convergence of Monte Carlo sampling, ensuring the efficiency of FDQMC. Instead, the access to Fock-state information benefits the ensemble selecting by imposing FRS on the Markov chain. The complexity of FDQMC is considerably reduced compared to the existing ensemble-restricted DQMC algorithms. This flexibility in accessing diverse ensembles broadens its applicability to non-Hubbard-like models, such as heavy-fermion systems with spin-fermion interactions, as discussed in this work.

The Fock-state sampling of FDQMC aligns closely with the cutting-edge quantum gas microscopy experiments with ultracold atoms in optical lattices\,\cite{gross2017quantum}. Therefore, FDQMC acts as a numerical emulator for fermionic quantum simulators, extending the range of conditions under which cold-atom experiments can be accurately calibrated. This is especially important for simulating higher-order correlations and entanglement-related properties, where minimizing statistical errors is necessary. With access to millions or even billions of sampled Fock-state samples, FDQMC further paves the way for the development of more sophisticated machine learning models. These models can extract in-depth insights beyond the rigorous quantum simulations, thereby expediting experimental discoveries\,\cite{bohrdt2019classifying}. 

\section*{Methods}
\vspace{-2mm}
\subsection*{Update strategy}
\vspace{-2mm}
For each update epoch, we alternately perform updates for the auxiliary field $x$ and the Fock state $\ket{\eta}$. The update algorithms leverage two different formulations of $Z_{x,\ket{\eta}}$. On one side, $Z_{x,\ket{\eta}}$ is a determinant
\begin{align}\label{eq:ZDeterminant}
    Z_{x,\ket{\eta}} = \det\mqty[P_{\ket{\eta}}^\T B_x P_{\ket{\eta}}]\,.
\end{align}
Here, $\qty[P_{\ket{\eta}}]_{ij} = \delta_{i,p_j}$ is the projection matrix of a Fock state $\ket{\eta}$, with $p_j$ denoting the site of the $j$-th particle. $B_x$ represents the single-particle propagator $\mathcal{T}e^{-\int_0^\beta H[x(\tau)]\dd{\tau}}$, associated with $x$. Eq.~\eqref{eq:ZDeterminant} shares the same mathematical structure used in the zero-temperature projective quantum Monte Carlo (PQMC)\,\cite{sugiyama1986auxiliary, sorella1989novel, white1989numerical, assaad2008world}. Hence, the $x$-direction update follows the same strategy as PQMC.

At the same time,  $Z_{x,\ket{\eta}}$ is proportional to the multi-point correlation under the auxiliary field configuration $x$ in the grand-canonical ensemble,
\begin{align}\label{eq:ZMultiPointCorr}
    Z_{x,\ket{\eta}} = \trace[ \rho_x] \expval{\prod_{i = 1}^{N_m} n_i(\eta)}_x\,.
\end{align}
In this formula, $n_i(\eta)$ denotes the electron (hole) density at site $i$, if the site is (is not) occupied in the Fock state $\ket{\eta}$ and $N_m$ represents the system size. To facilitate the rank-1 update for $\ket{\eta}$, we construct an auxiliary matrix
\begin{align}\label{eq:MmatConstruction}
    M_{\ket{\eta}} = \qty[\mqty(\eta_1 & & & \\ & \eta_2 & &\\ & & \ddots & \\ & & & \eta_{N_m}) - G^{(\rm gr)}_x]^{-1},
\end{align}
where $\eta_i$ denotes electron density of $\ket{\eta}$ at site $i$ and $G^{(\rm gr)}_x = (I + B_x)^{-1}$ is the equal-time Green's function under auxiliary field $x$ in a grand-canonical ensemble. The numerically stable inversion of Eq.~\eqref{eq:MmatConstruction} is presented in Supplementary Note 5. The acceptance ratio for a proposed a $\eta$-flip at site $i$, namely $\eta'_j = \eta_j + (-1)^{\eta_i}\delta_{ij}$, is
\begin{align}\label{eq: eta flip ratio}
    R_{\rm acc} = -1 - (-1)^{\eta_i}\qty[M_{\ket{\eta}}]_{ii}.
\end{align}
Upon acceptance of the flip, then $M_\eta$ is updated as
\begin{align}\label{eq: eta rank-1 update}
    \qty[M_{\ket{\eta'}}]_{jk} = \qty[M_{\ket{\eta}}]_{jk} + \frac{(-1)^{\eta_i}}{R_{\rm acc}}\qty[M_{\ket{\eta}}]_{ji}\qty[M_{\ket{\eta}}]_{ik}.
\end{align}
Each epoch of the Monte Carlo updates consists of first updating $x$ across all spatial and temporal sites, followed by proposing $\eta$-flips for each spatial site. The derivation of Eq.~\eqref{eq:MmatConstruction}, \eqref{eq: eta flip ratio}, and \eqref{eq: eta rank-1 update} is in Supplementary Note 6.

The computational complexities for constructing the initial $M_{\ket{\eta}}$  and for performing the iterations of $\eta$-flips both scale as $\order{N_m^3}$. With the Sherman-Morrison fast update\,\cite{white1989numerical}, the complexity of updating the auxiliary field $x$ is $\order{\beta N_m^3}$. Due to the imaginary time dimension, the cost for the latter overwhelms that of the $\eta$-upates. Therefore, the computational cost for FDQMC is only marginally more than that of the standard DQMC.

\subsection*{Fock-state restrictive updates}
\vspace{-2mm}
In the canonical ensemble, where the particle number is fixed, the Fock-state is updated by randomly swapping a particle at site $i$ with a hole at site $j$, as shown in Fig.\,\ref{fig:FRSschematic}\textbf{a}. The acceptance ratio for such a swap is given as
\begin{align}
\begin{aligned}
    &R^{\rm swap}_{\rm acc} = 1 + (-1)^{\eta_i}\qty[M_{\ket{\eta}}]_{ii} + (-1)^{\eta_j}\qty[M_{\ket{\eta}}]_{jj}\\
    &+ \mkern-2mu(-1)^{\eta_i+\eta_j}\qty(\qty[M_{\ket{\eta}}]_{ii}\mkern-1mu\qty[M_{\ket{\eta}}]_{jj} \mkern-3mu- \mkern-3mu\qty[M_{\ket{\eta}}]_{ij}\mkern-1mu\qty[M_{\ket{\eta}}]_{ji}).
\end{aligned}
\end{align}
Upon acceptance of the swap, $\ket{\eta}$ and $M_{\ket{\eta}}$ are updated by successively imposing single-site $\eta$-flips [i.e.~Eq.~\eqref{eq: eta rank-1 update}] at the sites of $i$ and $j$.

The complexity of widely used canonical DQMC algorithms\,\cite{ormand1994demonstration, rombouts1998accurate, gilbreth2015stabilizing}, which derive canonical ensemble properties by projecting from the grand-canonical ensemble using a Fourier projector, scales as $\order{\beta N_m^4}$ for observables beyond two-point correlations. In contrast, FDQMC avoids this additional overhead by directly sampling the canonical ensemble, keeping the complexity at $\order{\beta N_m^3}$. Meanwhile, truncation algorithms are applicable to FDQMC to further reduce the complexity of simulating dilute fermionic systems\,\cite{gilbreth2021reducing}.
\vspace{2mm}

\subsection*{Static and dynamical observables}
\vspace{-2mm}
The expectation value of general observable $O$ is evaluated by decomposing into the Green's function using Wick's theorem. The equal-time Green's function $[G_{x,\ket{\eta}}]_{ij}(\tau,\tau) = \expval{c_i(\tau)c_j^\dagger(\tau)}_{x,\ket{\eta}}$ for a specific $(x,\ket{\eta})$ is
\begin{equation}
    G_{x,\ket{\eta}}(\tau,\mkern-2mu\tau) \mkern-2mu= \mkern-2muI - B_{x}\mkern-2mu(\tau,\mkern-2mu0)P_{\ket{\eta}}\mkern-4mu\qty[\mkern-2muP_{\ket{\eta}}^\T B_{x}(\beta,\mkern-2mu0)P_{\ket{\eta}}\mkern-2mu]^{-1}\mkern-6muP_{\ket{\eta}}^\T B_{x}(\beta,\mkern-2mu\tau),
\end{equation}
where the single-particle propagator from $\tau_1$ to $\tau_2$ is $B_x(\tau_2,\tau_1) = \mathcal{T}e^{-\int_{\tau_1}^{\tau_2} H[x(\tau)]\dd{\tau}}$. Note that $G_{x,\ket{\eta}}$ differs from the grand-canonical $G^{\rm (gr)}_x$ with unspecified $\ket{\eta}$. For observables $O = \sum_{\ket{\eta}} w_{\ket{\eta}} \dyad{\eta}$ that are diagonal in real space, e.g., the charge and spin correlations, $\expval{O}_{x,\ket{\eta}}$ simplifies to $w_{\ket{\eta}}$. This simplification significantly reduces the computational cost for higher-order correlations by directly evaluating the sampled Fock states.

The single-particle spectrum is computed via $A(\vv k,\omega) = A_{+}(\vv k,\omega) + A_{-}(\vv k,\omega)$, where the particle-addition spectrum $A_{+}(\vv k,\omega)$ and the particle-removal spectrum $A_{-}(\vv k,\omega)$ are extracted from
\begin{align}
    \expval{c_{\vv ks}^\dagger(\tau)c_{\vv ks}(0)} &= \int A_{+}(\vv k,\omega) \ee^{-\tau\omega}\dd{\omega},\\
    \expval{c_{\vv ks}(\tau)c^\dagger_{\vv ks}(0)} &= \int A_{-}(\vv k,\omega) \ee^{\tau\omega}\dd{\omega},
\end{align}
separately, with $c_{\vv ks}^\dagger = (1/\sqrt{N})\sum_{\vv r}c_{\vv rs}^\dagger\ee^{\ii \vv k\vdot\vv r}$. The dynamic spin structure factor $S(\vv q, \omega)$ is obtained through
\begin{align}
    \expval{S_{\vv q}^{\rm z}(\tau)S_{\vv q}^{\rm z}(0)} = \int S(\vv q, \omega) \ee^{-\tau\omega}\dd{\omega}
\end{align}
with $S_{\vv q}^{\rm z} = (1/\sqrt{N})\sum_{\vv r}S_{\vv r}^{\rm z}\ee^{\ii \vv q\vdot\vv r}$. Both spectra in real frequency are calculated using the analytic continuation through the maximum entropy method\,\cite{jarrell1996bayesian}.

\section*{Data availability}
\vspace{-2mm}
The data supporting the findings of this study are available in the public repository Figshare at \href{https://doi.org/10.6084/m9.figshare.28146974}{https://doi.org/10.6084/m9.figshare.28146974}.

\section*{Code availability}
\vspace{-2mm}
The code is available upon request from the corresponding author.

\section*{Acknowledgements}
\vspace{-2mm}
We thank Immanuel Bloch, Annabelle Bohrdt, Thomas Chalopin, Fabian Grusdt, and Timon Hilker for their insightful discussions. This work is supported by the U.S. Department of Energy, Office of Science, Basic Energy Sciences, under Early Career Award No.~DE-SC0024524. The simulation used resources of the National Energy Research Scientific Computing Center (NERSC), a U.S. Department of Energy Office of Science User Facility located at Lawrence Berkeley National Laboratory, operated under Contract No.~DE-AC02-05CH11231 using NERSC award BES-ERCAP0031226.

\section*{Author Contributions}
S.D. developed FDQMC codes and performed calculations under the supervision of Y.W. S.L. assisted the data analysis. All authors contributed to writing the manuscript.

\section*{Competing interests}
The authors declare no competing interest.

\newpage
\onecolumngrid 
\vspace{8mm}
\begin{center}
    \large \textbf{References}
\end{center}
\vspace{-8mm}
\twocolumngrid 

\end{document}